\documentclass[a4paper]{jpconf}
\usepackage{graphicx}
\usepackage{xspace}
\usepackage{lineno}
\newcommand{\pbpb}       {\mbox{Pb--Pb}\xspace}

\begin{document}
%\linenumbers

\title{(W)hole new field: the new GEM Time Projection Chamber of ALICE}

\author{Dariusz Mi\'skowiec on behalf of the ALICE Collaboration}

\address{GSI Helmholtzzentrum f\"{u}r Schwerionenforschung, Planckstr. 1, 64291 Darmstadt, Germany}

\ead{d.miskowiec@gsi.de}

\begin{abstract}
The ALICE Time Projection Chamber has just been upgraded by replacing 
the readout chambers, which cover the two ends of the cylinder, with 
new detectors based on the Gas Electron Multiplier (GEM) technology. 
We report here on the related activities at GSI Darmstadt: 
GEM framing and chamber production, as well as the quality 
assurance accompanying both procedures. 
\end{abstract}

\section{Introduction}

\begin{figure}[b]
\hspace*{5mm}\includegraphics[width=0.92\textwidth]{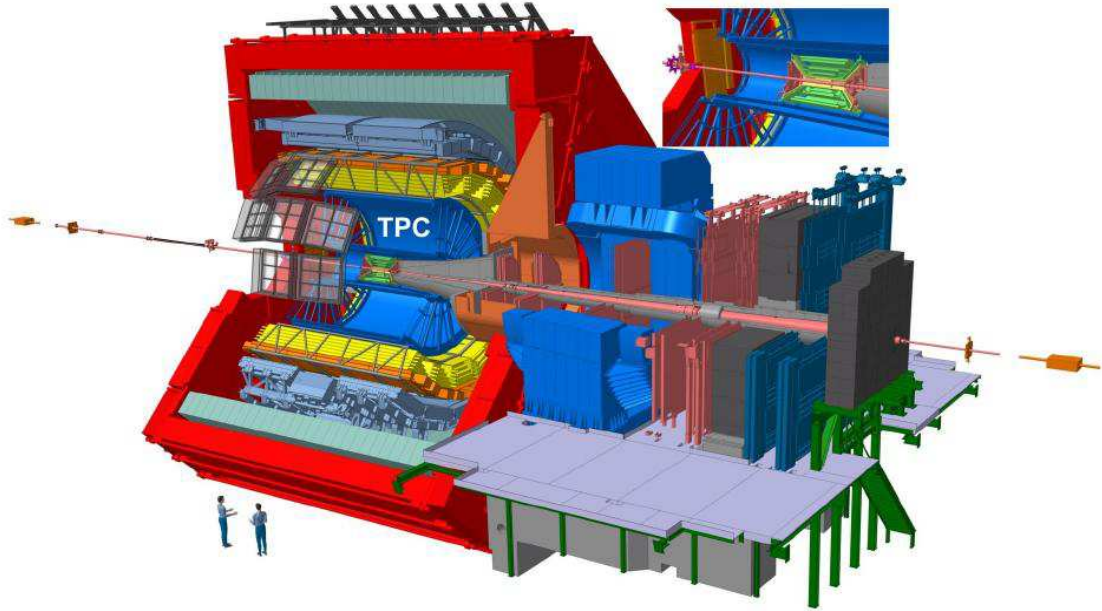}\hspace{2pc}%
\caption{\label{alice}The ALICE apparatus at the CERN LHC. The Time Projection 
Chamber is the main tracking and particle identification detector of
the ALICE central barrel.}
\end{figure}

The ALICE Time Projection Chamber (TPC) (Fig.~\ref{alice}) 
is the world's largest detector of its kind~\cite{alicetpc}. It served
as the main central tracking and particle-identification detector of
ALICE from 2009 till now (see Refs.~\cite{Aamodt:2008zz} and
\cite{Abelev:2014ffa} for the description of the experiment and its
performance). Its upgrade is an essential part of the experiment's
preparation for the LHC Run 3 starting in 2021, when the machine will
deliver \pbpb collisions at a rate of
50~kHz~\cite{ALICE:2014qrd,TheALICECollaboration:2015xke}.

\begin{figure}[h]
\hspace*{18mm}\rotatebox{90}{\includegraphics[height=0.77\textwidth]{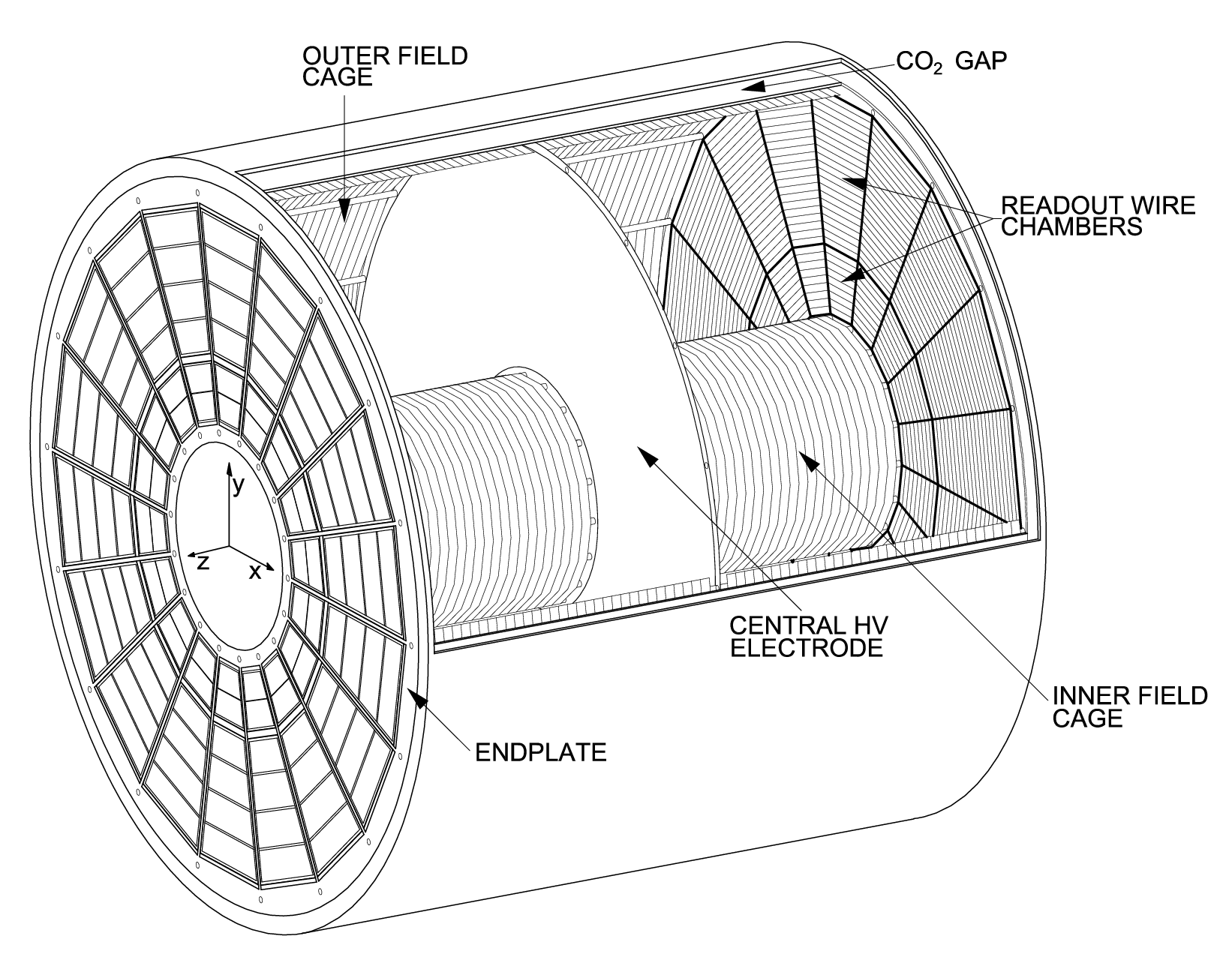}}
\caption{\label{tpc} The ALICE Time Projection Chamber. The ionization
  electrons drift parallel to the cylinder axis towards its ends,
  where they are amplified and produce signals in the readout
  chambers. With a segmentation of 20$^{\rm{o}}$ in azimuth and two
  sizes in radial direction, there are 36 inner and 36 outer readout
  chambers (IROCs and OROCs). These chambers have just been replaced
  by GEM-based chambers.  Half of the new OROCs was assembled and
  tested at GSI.}
\end{figure}

The principle of operation of the ALICE TPC as used in Runs 1 and 2 
is as follows. 
The TPC is filled with Ne-CO$_2$-N (90-10-5). Charged particles
traversing its drift volume ionize gas atoms.  An electric field of
400~V/cm makes the ionization electrons drift towards the readout
chambers (Fig.~\ref{tpc}) in which the charges are amplified and
produce measurable signals. The nominal gas amplification factor is
7000--8000.  The ions produced during the amplification slowly drift
towards the cathode and, if not stopped, would result in a space
charge within the active volume. The space charge would affect the
electron drift in the subsequent events. In order to avoid this, a
dedicated wire plane (gating grid) is located at the boundary between
the drift volume and the readout chambers. About 100~$\mu$s after
every trigger, after the last ionization electrons have arrived to the
readout chambers but before the first ions reach the drift volume,
the gating grid is made opaque by applying alternating voltages
to its wires. The gating grid is kept closed for 180~$\mu$s. This
method solves the problem of ion backflow, at the expense of
introducing a significant deadtime limiting the trigger rate to about
3~kHz.

The ongoing upgrade of the CERN accelerators will allow for \pbpb 
collision rates of 50~kHz from 2021 on. In order to fully make use 
of this, the readout chambers of the ALICE TPC have just been replaced 
by chambers of similar geometry but without the gating grid. Also 
the cathode and anode wire planes have been removed, the 
gas amplification is now performed sequentially in four layers of GEMs. Most of 
the ions are thus produced at the last GEM layer, and their way to the 
drift volume is made difficult by the presence of the first three 
GEM foils. With 8 potentials to tweak and with a somewhat reduced 
total gain factor, configurations can be found that result in an 
acceptable ion backflow ($<$1\%) while preserving the energy resolution 
(12\% for 5.9 keV x-rays). 
The parameters of wire and GEM chambers are compared in Table~\ref{tab1}. 
After eliminating the gating grid and the related dead time, the plan 
is to run the upgraded TPC in a continuous -- rather then triggered -- 
readout mode, thus recording all \pbpb interactions offered by the LHC. 
\begin{center}
\begin{table}[h]
\caption{\label{tab1}Comparison of wire and GEM chambers.}
%\footnotesize\rm
\centering
\begin{tabular}{lccc}
\br
& \multicolumn{2}{c}{wire chamber} & GEM chamber\\
& grid open & grid closed &\\
\mr
gain & 8000 & 0 & 2000\\
ion backflow & 0.13 & $<0.0001$ & $<$0.01 \\
\br
\end{tabular}
\end{table}
\end{center}

\section{GEM-chamber production}
\begin{figure}[b]
\vspace*{-3mm}
\hspace*{3mm}
\includegraphics[width=0.95\textwidth, clip, trim=0mm 0mm 0mm 10mm]{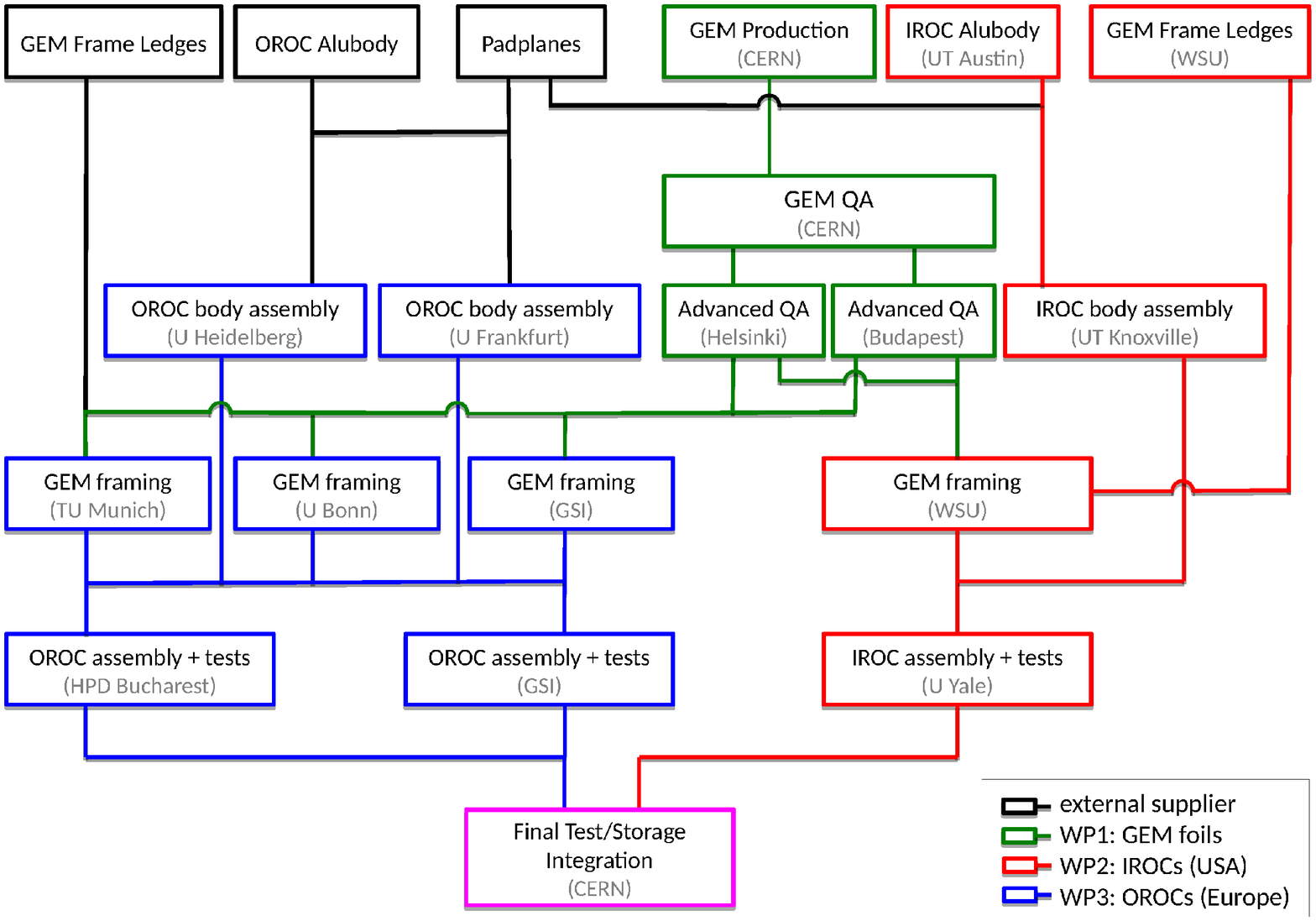}
\caption{\label{scheme}Scheme of GEM-chamber production project. The
  GSI team framed the largest GEM foils and assembled and tested 50\%
  of the Outer Readout Chambers (OROCs).}
\end{figure}

The ALICE TPC Upgrade collaboration consists of teams from 52 institutions. 
The chamber production was highly decentralized (Fig.~\ref{scheme}). 
The GSI team joined the detector building activities in 2015, getting 
involved in GEM framing as well as assembly and tests of OROCs. 
The relevant steps of the production scheme are described below. 

\begin{figure}[t]
\hspace*{10mm}\includegraphics[width=0.85\textwidth]{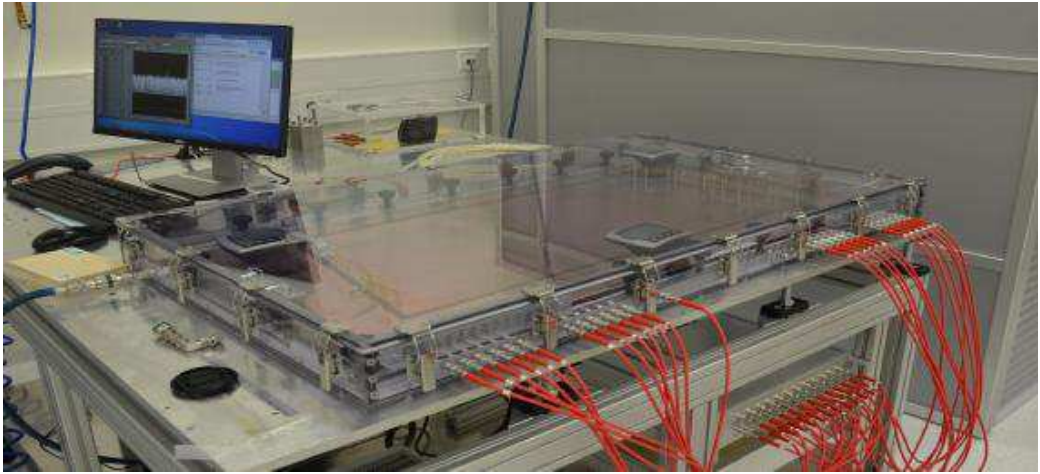}
\caption{\label{gemtest}Measurement of the GEM dark current. The box 
is flushed with nitrogen, 500 V are applied and the currents are 
measured separately for all segments.} 
\end{figure}

\begin{figure}[b]
\vspace*{-3mm}
\hspace{15mm}\includegraphics[width=0.8\textwidth]{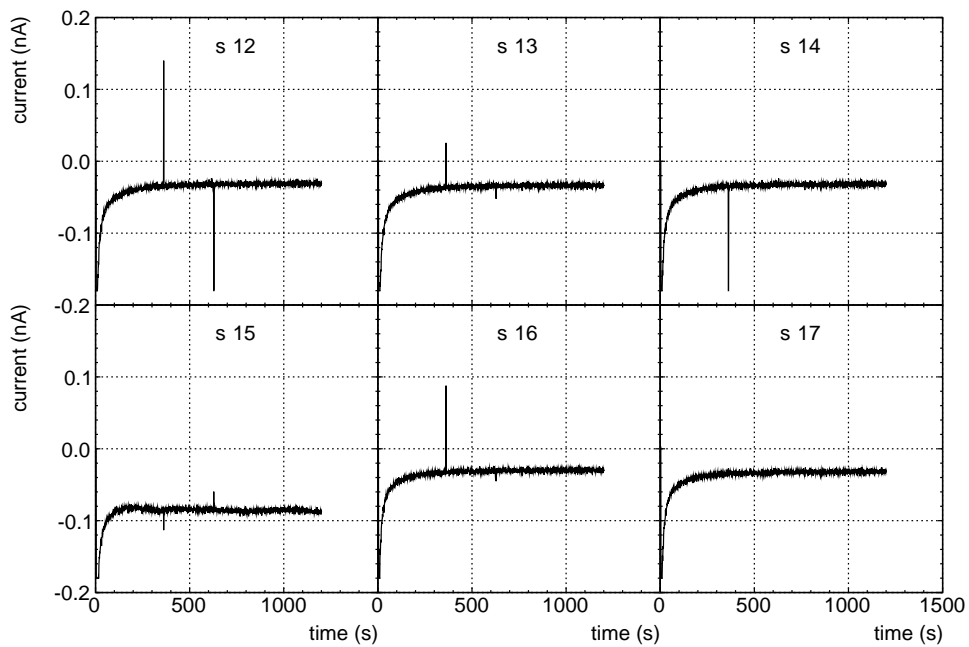}
\caption{\label{gemcurrents}Example of a pathological GEM. Segment
15 (lower left panel) has a significantly higher current than others.}
\end{figure}

A basic test of a GEM, as performed in particular before and after
every transport, consisted in measuring its dark current. For this, a
dry GEM (a few days spent at a relative humidity $<$1\%) was placed in
a plexiglass box equipped with spring-loaded pin contacts, and flushed
with nitrogen (Fig.~\ref{gemtest}). A voltage of 500 V was applied to
all segments (20--24, depending on GEM size) and the respective
currents were monitored.
Typically, the currents stabilized at a level of 10--50~pA, depending
on the hole pitch of a GEM. The currents of individual segments within
a GEM normally agreed within 10\%.  An excessive current of a segment
meant a shorted GEM.  Increased current (see example in
Fig.~\ref{gemcurrents})
may come from a defect/contamination and was considered a potential
danger.  Shorted and contaminated GEMs had a fair chance of recovery
when sent back to their production site for recleaning.

\begin{figure}[t]
\hspace{20mm}\includegraphics[width=0.75\textwidth]{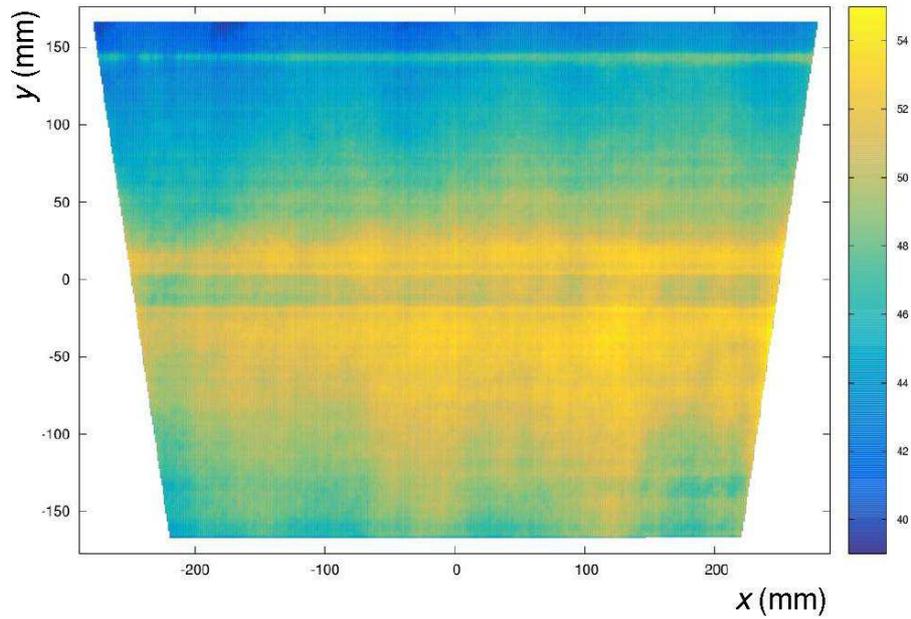}
\caption{\label{holesize}Example of a hole size map, extracted from 
optical survey of a GEM. } 
\end{figure}

An advanced quality assurance procedure, performed once for each GEM,
consisted in a long-term (at least 5 hours) dark-current measurement
and an optical survey. During the latter, microscope photographs were
taken of the entire GEM. The pictures were stitched together and
analyzed for defects and hole-size nonuniformities
(Fig.~\ref{holesize}).  
The gain of a single GEM appeared to be clearly
anticorrelated with the diameter of the holes.  This correlation, 
however, was practically lost in a stack of four GEMs.

\begin{figure}[b]
\hspace{25mm}\includegraphics[width=0.64\textwidth]{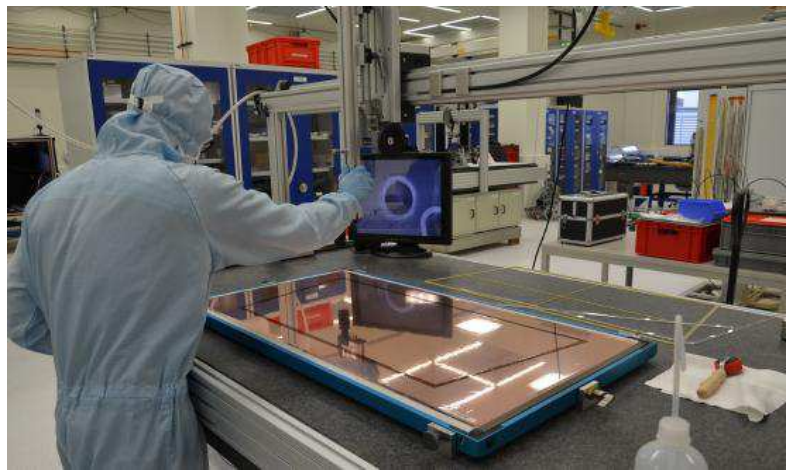}
\caption{\label{alignment}The GEM foil is being aligned to the epoxy-covered 
frame which is underneath it. } 
\end{figure}

Framing a GEM consisted in stretching it and positioning 0.5 mm above a
rigid frame made of G11 with a thin layer of epoxy glue on it.  After
a careful alignment (Fig.~\ref{alignment}), 
the GEM foil was pressed to the frame and covered with a plexiglass
hood. The box was flushed with nitrogen and left overnight. 

The complete information about the 923 GEM foils of the project was
stored in a dedicated database.  With an intuitive user interface and
numerous analysis macros, the database was a valuable tool in
steering and monitoring the production process.

\begin{figure}[t]
\hspace{25mm}\includegraphics[width=0.7\textwidth]{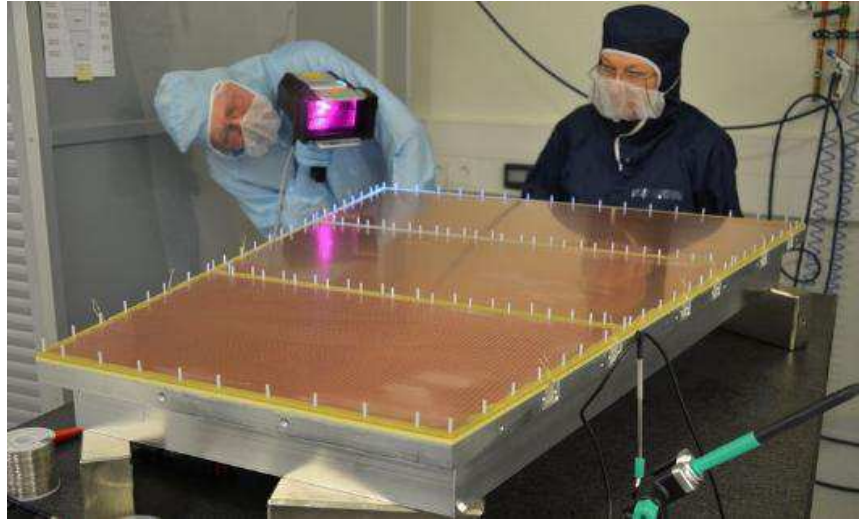}
\caption{\label{body}Chamber body before GEM installation. Readout pad
  planes of the three stacks and nylon bolts for GEM mounting are
  visible. The pad planes are inspected under UV light.}
\end{figure}

\begin{figure}[b]
\hspace{25mm}\includegraphics[width=0.72\textwidth]{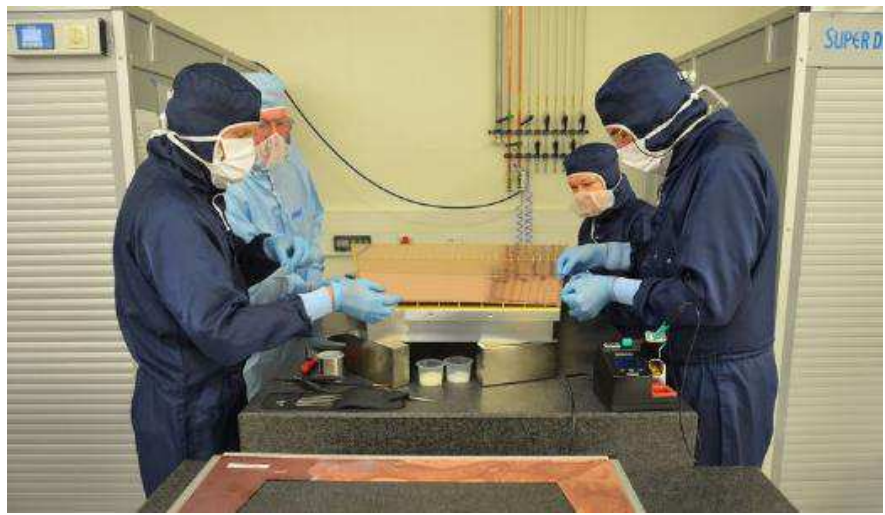}
\caption{\label{adding}Chamber assembly. A GEM is being added to the 
stack. After adding the fourth -- last -- GEM foil, the nylon nuts holding 
stacks together are tightened with a torque of 6~Ncm.} 
\end{figure}

The OROC chamber assembly consisted in adding three stacks of 4 GEMs
each to the chamber bodies (Figs.~\ref{body}, \ref{adding}). 
The bodies were equipped with pad planes and connection wires  
bringing HV into the chambers and the signals out of it. The GEMs 
were attached to them with densely spaced nylon bolts
(Fig.~\ref{body}),
ensuring a reasonable level of tension. The GEMs were trimmed to
trapezoidal shape and mounted one by one. The HV connections were
soldered to both sides of each GEM. As anticipated, the size, shape,
and precise location of the soldering points were of high importance
for the HV stability of the chambers.

\section{GEM-chamber quality assurance}

\begin{figure}[h]
\hspace{12mm}\includegraphics[width=0.9\textwidth]{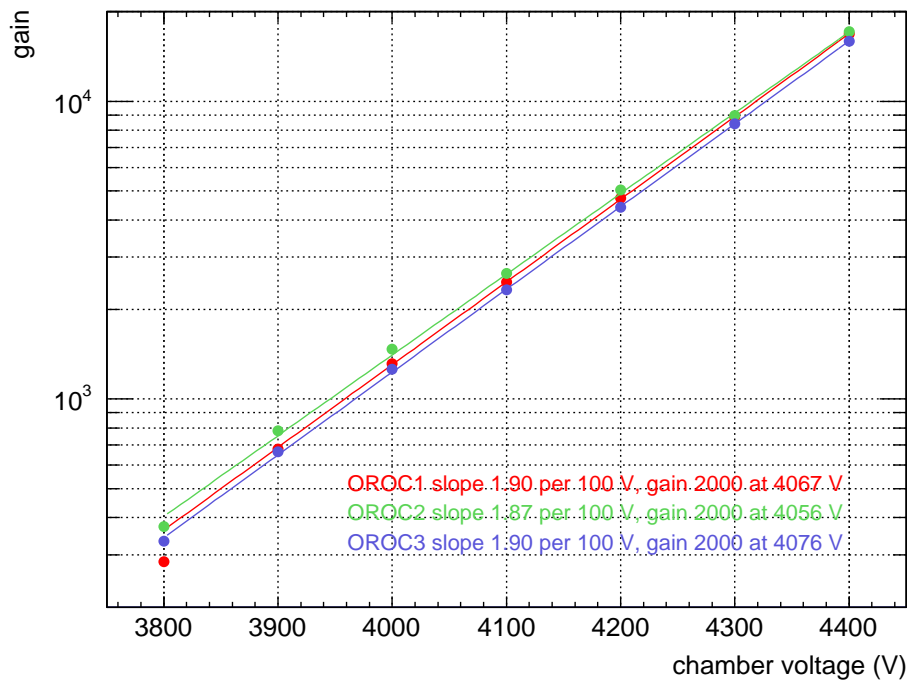}
\caption{\label{gain} Gain curves of the three GEM stacks of one
  particular chamber (OROC/23). As expected, the gain increases
  exponentially with the applied voltage.  The difference between
  stacks can be compensated by applying slightly different
  voltages. The nominal gain is 2000.}
\end{figure}

The chambers were subject to several tests at the assembly site. For
this, they were enclosed in a test box filled with the nominal gas
mixture, with a low-mass entrance windows and a 13~mm drift gap with
400~V/cm before the first GEM. The gain of the chamber, for each of
the three GEM stacks separately, was measured by irradiating it with
5.9 keV x-rays from a $^{55}$Fe source. The photons that reach the
drift gap ionize gas atoms, producing on average a primary charge of
166 electrons. The absolute gain factor is calculated by dividing the
pad current by the x-ray rate (measured using a scaler) and the
primary charge. An example of the obtained gain curves is shown in
Fig.~\ref{gain}.

\begin{figure}[p]
\hspace{15mm}\includegraphics[width=0.78\textwidth]{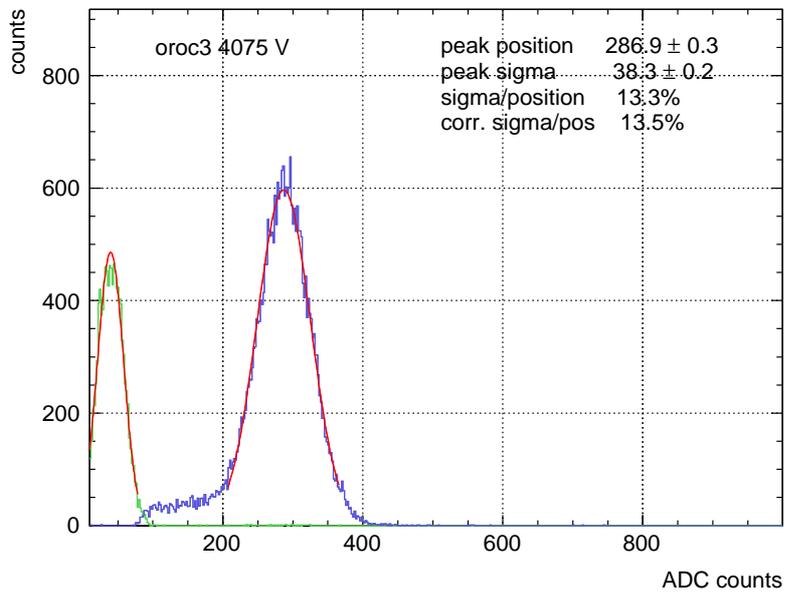}
\caption{\label{dedx} X-ray spectrum of $^{55}$Fe, measured with the
  third stack of chamber OROC/23. The relative width of the 5.9 keV
  peak, corrected for the position and width of the pedestal (smaller
  peak on the left side), is the measure of the energy resolution.}
\end{figure}

The charge induced on the pads is proportional to the primary ionization 
and thus to the energy deposited by the traversing charged particle. 
The energy resolution is checked by recording the $^{55}$Fe x-ray spectrum 
and fitting the 5.9~keV peak (Fig.~\ref{dedx}).  
The test is performed at a voltage corresponding to a gain of 2000.
The relative width of the peak, calculated taking into account the
position and the width of the pedestal, was typically around 12-14\%,
usually slightly above the limit of 12\%. As the energy resolution and the ion
backflow can be traded against each other by changing the GEM
voltages, and as we notoriously observed an ion backflow significantly
better than the limit, the poor values of the energy resolution were
not considered to be a problem.

\begin{figure}[p]
\hspace{15mm}\includegraphics[width=0.74\textwidth]{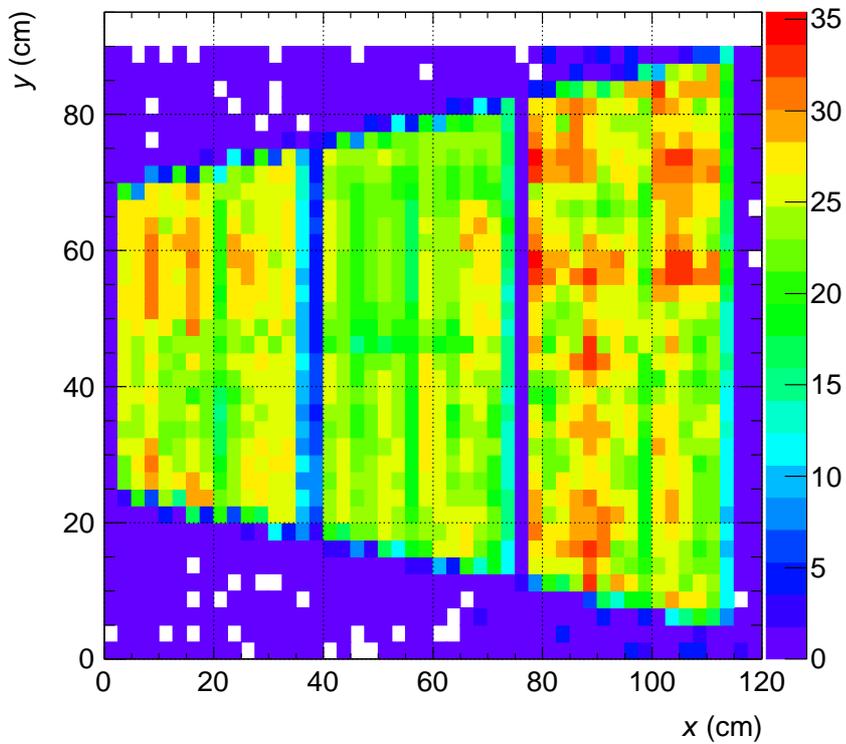}
\caption{\label{gainmap} The pad current (in nA) as a function of the
  position of the x-ray tube for chamber OROC/23. The gain uniformity
  is 11\% (standard deviation), well below the requirement of 20\%.}
\end{figure}

The gain homogeneity was evaluated by shifting a collimated x-ray tube
over the chamber surface and monitoring the anode current (sum of all
pads). This test was performed with the three stacks kept at a voltage
corresponding to a gain of 2000. An example of the resulting current
map is shown in Fig.~\ref{gainmap}.

\begin{figure}[h]
\hspace{15mm}\includegraphics[width=0.75\textwidth]{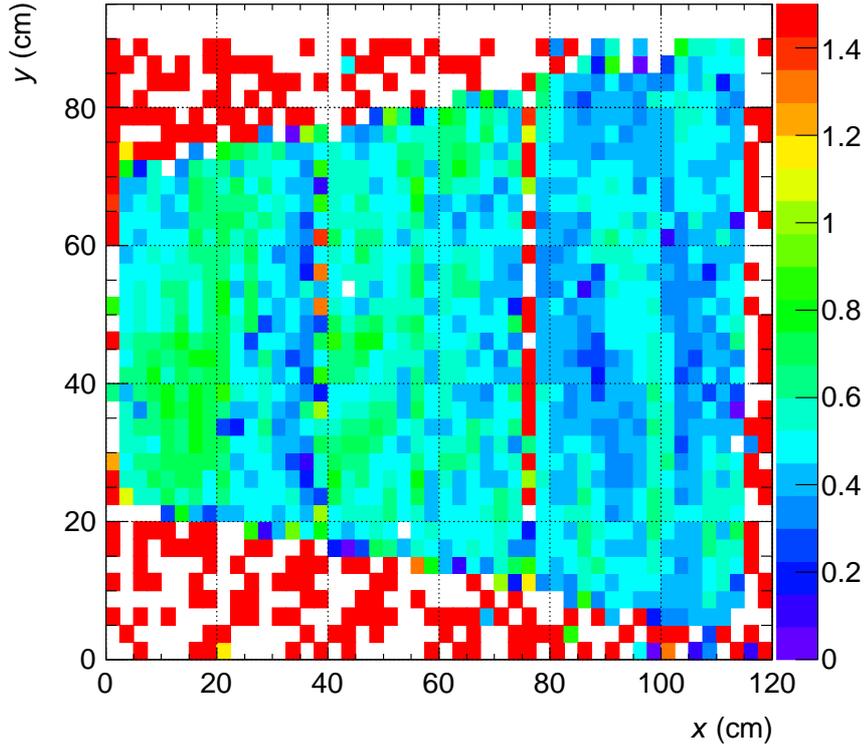}
\caption{\label{ibfmap} The ion backflow factor (in percent) as a
  function of the position of the x-ray tube for chamber OROC/23. The
  IBF factor was calculated as the ratio between the cathode and pad
  currents. Its mean value is 0.5\%, well below the required limit of
  1\%.}
\end{figure}

The ion backflow (IBF) measurement is performed in parallel with the
gain uniformity test. For this purpose, the cathode current is
recorded along with the pad one. The cathode current is caused by
ions traversing the drift volume and reaching the cathode. The IBF
factor is calculated as the ratio between the cathode and the pad 
currents. An example map is shown in Fig.~\ref{ibfmap}.

The stability of each chamber was checked by illuminating it
semi-uniformly with two strong x-ray tubes. The level of irradiation
was such that the pad current was 10 nA/cm$^2$ and the test duration
was 6 hours. The pad and cathode currents were monitored. After the
test, the health of all GEMs was checked by measuring their dark
currents.

Upon completion of the tests, each chamber was enclosed in its
transportation box. The gas tightness was checked and the boxes were
flushed with nitrogen. After the last check of the GEM integrity,
performed by applying 250 V to each GEM and measuring its dark
current, the chambers were shipped to CERN.

\section{GEM chamber installation}
\begin{figure}[t]
\hspace{15mm}\includegraphics[width=0.74\textwidth]{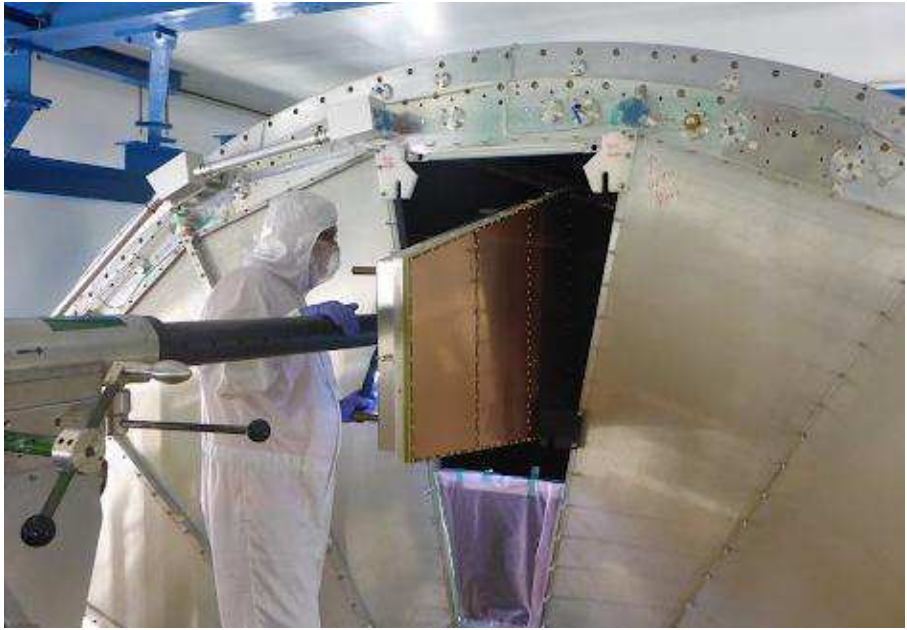}
\caption{\label{installation} Chamber replacement in the ALICE
  TPC. The operation took six months and was performed in a dedicated
  cleanroom at CERN.}
\end{figure}

In the beginning of 2019, the ALICE TPC was extracted and brought to a
dedicated cleanroom on the surface. The replacement of the chambers
started in April and was completed in September. Installation of a
chamber requires inserting it into the TPC, turning, and pressing from
inside against the backplate (Fig.~\ref{installation}). The removal
of wire chambers and the installation of GEM chambers each proceeded at a
pace of two sectors per day. Replacing all 72 chambers took six months. 

\section{Concluding remarks}
The involvement of the GSI group in the construction of the GEM 
chambers for the ALICE TPC upgrade started in 2015 with planning, 
ordering equipment, and defining procedures. The peak production 
took place in 2017--2018. With the new GEM technology and large 
detector sizes, the production encountered several surprises and 
some procedures had to be adjusted correspondingly. In all 
production steps, strong emphasis was put on quality assurance. 

The works were performed in the GSI detector laboratory. It was 
a privilege and pleasure to use its excellent infrastructure. 
Our hosts, the detlab colleagues, were welcoming and helped to 
quickly overcome some of the encountered problems. We appreciate 
their purely science- and curiosity-driven motivation and 
gratefully acknowledge their contribution to our project. 

By the end of 2019, the time of submission of these proceedings, the 
chambers have been installed in the TPC, and the TPC commissioning 
has just started. We are all looking forward to seeing the upgraded 
TPC performing beautifully in the coming LHC Run 3.


\section*{References}
\begin{thebibliography}{9}
\bibitem{alicetpc} Alme J {\it et al} 2010 
The ALICE TPC, a large 3-dimensional tracking device with fast readout for 
ultra-high multiplicity events
{\it Nucl. Instrum. Meth.} {\bf A622} 316-367
\bibitem{Aamodt:2008zz}
  Aamodt K {\it et al.} [ALICE Collaboration] 2008 
  The ALICE experiment at the CERN LHC 
  {\it JINST} {\bf 3} S08002
\bibitem{Abelev:2014ffa}
  Abelev B {\it et al.} [ALICE Collaboration] 2014 
  Performance of the ALICE Experiment at the CERN LHC
  {\it Int.\ J.\ Mod.\ Phys.}\ A {\bf 29} 1430044
\bibitem{ALICE:2014qrd}
  ALICE Collaboration 2013 
  Upgrade of the ALICE Time Projection Chamber 
  {\it CERN-LHCC-2013-020, ALICE-TDR-016}
\bibitem{TheALICECollaboration:2015xke}
  ALICE Collaboration 2015 
  Addendum to the Technical Design Report for the Upgrade of the ALICE Time Projection Chamber
  {\it CERN-LHCC-2015-002, ALICE-TDR-016-ADD-1}
\end{thebibliography}
\end{document}